\documentstyle[11pt]{article}
\newcommand{\Sekt}[1]{\section{#1} \setcounter{equation}{0}}
\newcommand{\BEQ}{\begin{equation} }
\newcommand{\EEQ}{\end{equation} }
\newcommand{\BEA}{\begin{eqnarray} }
\newcommand{\EEA}{\end{eqnarray} }
\newcommand{\BEAn}{\begin{eqnarray*} }
\newcommand{\EEAn}{\end{eqnarray*} }
\newcommand{\Balk}[1]{\rule[-#1em]{0em}{#1em}}

\newcommand{\ON}{\mbox{O$(N)$}}

\newcommand{\rf}[1]{\mbox{(\ref{#1})}}

\newcommand{\ds}{\displaystyle}
\newcommand{\cZ}{\mbox{${\cal Z}$}}
\newcommand{\cN}{\mbox{${\cal N}$}}
\newcommand{\cO}{\mbox{${\cal O}$}}

\newcommand{\cat}{catastrophe\ }
\newcommand{\cats}{catastrophes\ }
\newcommand{\hlf}{\frac{1}{2}}
\newcommand{\Nlf}{\frac{N}{2}}
\newcommand{\twrd}{\frac{2}{3}}
\newcommand{\trd}{\frac{1}{3}}
\newcommand{\tlf}{\frac{3}{2}}

\title{Double Scaling Limits and Catastrophes of the zerodimensional O$(N)$
Vector Sigma Model: The A-Series\\[0.6cm]
}
\author{W. R{\"u}hl\thanks{e-mail: ruehl@physik.uni-kl.de}, \\
        Fachbereich Physik, Universit\"at  Kaiserslautern,\\
        D-67653 Kaiserslautern, Germany}
\date{\LaTeX-version of preprint KL-TH-94/21\\September 1994}

\begin{document}
\input epsf                       
\renewcommand{\theequation}{\arabic{section}.\arabic{equation}}
\maketitle
\begin{abstract} We evaluate the partition functions in the neighbourhood of
catastrophes by saddle point
integration and express them in terms of generalized Airy functions.
\end{abstract}
\Sekt{Introduction}
The vector models have attracted interest as one-dimensional quantum gravity
theories just as the matrix models are interpreted as representing two
dimensional quantum gravity theories. A connection with polymer models has
also been emphasized from the outset \cite{And}. Their double scaling limit
has been studied with the usual renormalization group and $1/N$ expansion
methods. In the zero dimensional case the beta function and the free energy
have been calculated for large $N$ exactly.

We will show that the double scaling limits for this most elementary model can
be calculated exactly, i.e. asymptotically to any order in a
$N^{-\frac{1}{m+1}}$-expansion. Catastrophes are singularities of
differentiable maps \cite{Arnold1,Arnold2} and by diffeomorphisms can be
transformed
to canonical forms. We will study such canonical forms only. It does not make
sense to reshape these canonical forms by application of diffeomorphisms.

There are elementary and nonelementary catastrophes. The elementary ones are
ordered into $A$,
$D$
and $E$ cases \cite{Arnold1}. Whereas vector models with one vector field can
exhibit only A-series \cats $(A_m,\;m\in I\!\!N)$, models with two vector
fields can also posses $D$ or $E$ series \cats. The nonelementary
catastrophes show up also first in two-field models.

Application of \cat theory to zero dimensional O$(N)$ sigma models
leads to a wealth of useful information which can be used as a guideline for
studies of more complicated models. The diffeomorphisms which are basic in
\cat theory replace the 'reparametrisations in coupling constant space' ;
those properties of \cats which hold true independently of
diffeomorphisms (characterize cosets of the diffeomorphism group) map onto
universal features of phase transitions.
This is of course not new. Nevertheless our analysis will lead to some new
insights.

We define the zero dimensional \ON vector models by
\BEA \label{eq:partition}
\cZ_N(g)&=& \int\,d\Phi\,\exp\Big\{ -N\,g(\Phi)\Big\},\qquad
d\Phi \;:\; \mbox{Lesbeque measure}, \quad\Phi\in I\!\!R_{I\!\!N}
\EEA
where $g(\Phi)$ is \ON\- invariant and has the asymptotic expansion
\BEA
g(\Phi)&\simeq&\hlf\sum\limits_{k=1}^\infty\,\frac{g_k}{k}\,
\Big(\Phi^2\Big)^k
\EEA
Constraints on the $\{g_k\}$ lead to catastrophes that dominate the large $N$
behaviour of the partition function \rf{eq:partition} through saddle point
expansions. Main ingredients in these expansions are (real nonoscillating)
generalized Airy functions.

The free energy is defined by
\BEA
\label{eq:Fdef}F_N(g)&=&-\frac{1}{N}\log \cZ_N(g)\,-\,\hlf\log
\frac{Ng_1}{2\pi}
\EEA
where this normalization is such that
\BEA
F_N(g)\bigg|_{g_k=0\;\forall k\ge 2} &=&0
\EEA
The simplest \cat is of $A_1$ type (or Morse or Gaussian). We set
\BEA
&g_1\,=\,1,\quad g_2\,\ge\,0,\quad g_k\,=\,0\;\forall k\ge 3&
\EEA
then $\cZ_N$ can be expanded into a $1/N$ expansion around a Gaussian saddle
point
\BEA
F_N(g_2)\bigg|_{N=\infty}&=&\hlf \log \frac{1+\sqrt{\Delta}}{2}\,+\,
\hlf\frac{1}{1+\sqrt{\Delta}}\,-\,\frac{1}{4}
\EEA
with
\BEA \label{eq:Deltad}
\Delta&=&1\,+\,4g_2
\EEA
This function satisfies the (large $N$) Callan-Symanzik equation.
\BEA
\left(N\frac{\partial}{\partial N} - \beta(g_2)\frac{\partial}{\partial g_2}
+\gamma(g_2)\right)\,F_N(g_2)&=& R_N(g_2)
\EEA
Assume we know that
\BEA
\gamma(g_2)&=&1
\EEA
for the free energy. We can continue $F_N(g_2)\Big|_{N=\infty}$ analytically
off the
positive real axis till the neighbourhood of
\BEA
g_*&=& -\frac{1}{4}
\EEA
where $F_N(g_2)\Big|_{N=\infty}$ has a branch point in the variable $\Delta$
\BEA
F_N(g_2)\bigg|_{N=\infty}&=& g(\Delta)\,+\,\Delta^\tlf\,h(\Delta) \nonumber\\
\bigg( h(0)&=&-\frac{1}{3}\bigg )
\EEA
both functions $g$, $h$ being analytic.

Now the singular part is defined
to satisfy the homogeneous Callan-Symanzik (renormalization group or RG)
equation. It follows
\BEA
\beta(g_2)^{-1}&=& \frac{\partial}{\partial g_2}\log\Big(\Delta^{\tlf}
h(\Delta)\Big)\\
\beta(g_2)&=& \twrd(g_2-g_*) + {\cal O}\Big((g_2-g_*)^2\Big)
\EEA
and
\BEA
R_N(g_2)&=& \Big(-\beta(g_2)\,\frac{\partial}{\partial g_2} +1\Big)\,g(\Delta)
\EEA
The results on $\beta$, $F_n$, $R_N$ given in \cite{Hig} are thus reproduced.

In the neighbourhood $g_2\simeq g_*$ it is guessed that the singular part of
$F_N(g_2)_{\mbox{sing}}$ (which at $N=\infty$ is equal
$\Delta^{\tlf}h(\Delta)$) possesses the series expansion
\BEA \label{eq:Fguess}
F_N(g_2)_{\mbox{sing}} &=& \sum\limits_{h=0}^\infty\,a_h\,
N^{-h}\,(g_2-g_*)^{2-\gamma_0-\gamma_1 h}
\EEA
This is based on the fact that each term of \rf{eq:Fguess} satisfies the RG
equation when
\BEA
&\gamma_1\,=\,\frac{1}{\beta^\prime (g_*)}\,=\,\tlf , \qquad
\gamma_0\,=\,2-\frac{\gamma (g_*)}{\beta^\prime
(g_*)}\,=\,\hlf &
\EEA
However: this argument is too simple, since the RG equation does not determine
$h$ to be an integer. If we apply derivation of $h$ to any term in
\rf{eq:Fguess} we obtain also a permitted contribution
\BEA
&a^\prime_h\,N^{-h}\,(g_2-g_*)^{2-\gamma_0-\gamma_1 h}\,\log\Big[ N
(g_2-g_*)^{\gamma_1}\Big]&
\EEA
In fact such term for $h=1$ exists.

The $g_2$ parameter defines a curve of $A_1$ \cats which at $g_2=g_*$
ends in a $A_2$ catastrophe. But at this \cat and in a neighbourhood we
can derive $F_N(g_2)_{\mbox{sing}}$ directly (all of $F_N(g_2)$ in fact). The
result (Section 3) is
\BEA
F_N(g_2)_{\mbox{sing}} &=& -\frac{1}{N}\log \mbox{Bi}(\zeta)\\
\zeta&=&\Big(\Nlf\Big)^{\twrd}(g_2-g_*)
\EEA
and Bi is the usual Airy function. Moreover the expansion \rf{eq:Fguess} is
asymptotic only and
\BEA \label{eq:a1eq}
a_1^\prime&=&+\frac{1}{6}
\EEA

In section 4 we deform $A_2$ catastrophes by $g_3$. The aim is to render the
partition function for canonical $A_2$ convergent. A similiar problem with
convergence appears for all $A_m$,
$m$ even, and can be solved the same way. In section 5 we discuss the canonical
$A_m$ \cats for general $m$
and in section 6 we derive the corresponding $\beta$-functions.
\Sekt{A pedagogical example of mathematical interest}
The confluent hypergeometric function
\BEA
_1F_1(\alpha;\gamma;z)&=& \sum_{n=0}^\infty\,
\frac{(\alpha)_n}{(\gamma)_n}\,\frac{z^n}{n!},\qquad
(\alpha)_n\;=\;\frac{\Gamma(\alpha+n)}{\Gamma(\alpha)}
\EEA
is a well studied transcendental function known in mathematical physics since
more than hundred years. Nevertheless the asymptotic behaviour in the limit
\BEA
\label{eq:asyp1} &&\alpha\to\infty,\qquad z\to\infty,\qquad \gamma
\quad\mbox{fixed}\\
\label{eq:asyp2}&&\frac{\alpha}{z}\;=\;\xi,\\
&&\xi\to -\frac{1}{4},\quad\mbox{so that}\quad
(1+4\xi)z^{\twrd}\quad\mbox{is again fixed}
\EEA
is not covered in the textbook literature (see e.g. Luke's otherwise extremly
useful treatise \cite{Luke}). This is a typical 'double scaling limit'. The
standard approach is a saddle point integration technique.

The integral representation for $_1F_1$ is (\cite{Grad}, 9.211.2)
\BEA \label{eq:intdefF}
_1F_1(\alpha;\gamma;z)&=&
\ds\frac{\Gamma(\gamma)}{\Gamma(\alpha)\Gamma(\gamma-\alpha)}\int_0^1 dt t^{-1}
(1-t)^{\gamma-1}\exp\Big\{ zt+\alpha\log t-\alpha\log (1-t)\Big\}
\EEA
for $\Re (\alpha) > 0$, $\Re (\gamma-\alpha)>0$. Though the asymptotic region
\rf{eq:asyp1} lies outside the convergence domain of \rf{eq:intdefF}, the
saddle point expansion we shall derive remains valid. We consider only the
contribution of the saddle point and not those of the boundaries. In physical
applications one must be more careful. The relevant contribution is always the
one which dominates the asymptotic behaviour.

Define using \rf{eq:asyp2}
\BEA
f(t) &=& t+\xi\log\frac{t}{1-t}
\EEA
as 'phase function'. There are two extrema of $f(t)$ at $t_{\pm}$ if
$\xi>-1/4$ and none if $\xi<-1/4$
\BEA
\label{eq:tpm}t_{\pm}&=& \hlf \Big( 1\pm \sqrt{\Delta}\Big)\\
\Delta &=& 1+4\xi
\EEA
For $\Delta=0$ we obtain a point of inflexion at
\BEA
t_0&=&\hlf
\EEA
We expand $f(t)$ around $t_0$
\BEA
\label{eq:ftaylor}f(t)&=& f(t_0)+(t-t_0) f^\prime(t_0)+\frac{1}{6}(t-t_0)^3
f^{\prime\prime\prime}(t_0) +{\cal O}\Big((t-t_0)^4\Big)
\EEA
with
\BEA
&&f(t_0)\,=\,\hlf,\qquad f^\prime(t_0)\,=\,\Delta,\qquad
f^{\prime\prime\prime}(t_0)\,=\,32\xi
\EEA
where we may approximate
\BEA
f^{\prime\prime\prime}(t_0)&=&-8
\EEA
Now we scale the integration variable so that
\BEA
t-t_0 &=&\lambda\,\eta\\
z\,\frac{1}{3!}\,f^{\prime\prime\prime}(t_0)\lambda^3 &=& -\frac{1}{3}\\
\lambda &=& (4z)^{-\frac{\lambda}{3}}
\EEA
Thus the leading part of $_1F_1$ coming from this saddle point is
\BEA
&\label{eq:leading1F1}\ds\frac{\Gamma(\gamma)}{\Gamma(\alpha)\Gamma(\gamma-\alpha)}e^{zf(t_0)}
t_0^{-1} (1-t_0)^{\gamma-1} (4z)^{-\frac{1}{3}}\,\Phi(\zeta)&
\EEA
where $\Phi(\zeta)$ is of greatest interest for us
\BEA
\label{eq:Phintdef}\Phi(\zeta)&=& \int_C d\eta
e^{\zeta\eta-\frac{1}{3}\eta^3}\\
\zeta&=& 4^{-\frac{1}{3}} z^{\twrd}\Delta
\EEA
As a rule we use the average over two 'least deformed real axis' contours on
which
\rf{eq:Phintdef} converges.
Define
\BEA
r\in Q:&& C_r\,=\,\mbox{contour from zero to $e^{2\pi i r}\infty$ along a ray}
\EEA
Then in \rf{eq:Phintdef} we set
\BEA
\label{eq:Cchoice}C&=& C_0-\hlf (C_{\frac{1}{3}}+C_{-\frac{1}{3}})
\EEA
With the ray integrals
\BEA
R_r^{(3)}(\zeta)&=&\int_{C_r} d\eta e^{\zeta\eta-\frac{1}{3}\eta^3}
\EEA
we get
\BEA
\Phi(\zeta)&=& R_0^{(3)}(\zeta)-\hlf\Big(R_{\frac{1}{3}}^{(3)}(\zeta)
+R_{-\frac{1}{3}}^{(3)}(\zeta)\Big)
\EEA
Using small $\zeta$ expansions to identify functions we find (\cite{Abram},
10.4.3)
\BEA
\Phi(\zeta)&=& \pi\,\mbox{Bi}(\zeta)\\
\label{eq:wefind}\frac{-i}{2}\Big(R_{\frac{1}{3}}^{(3)}(\zeta)
-R_{-\frac{1}{3}}^{(3)}(\zeta)\Big) &=& \pi\,\mbox{Ai}(\zeta)
\EEA
{}From (\cite{Abram}, figs 10.6 and 10.7) we see that
\BEA
&\mbox{Bi}(\zeta_0)\,=\,0,\qquad \zeta_0\,=\,-1.173&
\EEA
and Bi$(\zeta)$ oscillates for $\zeta<\zeta_0$ and is positive for
$\zeta>\zeta_0$. The function Ai$(\zeta)$ oscillates everywhere. This justifies
the choice of of the contour $C$ \rf{eq:Cchoice} for $\zeta>\zeta_0$.

The Airy functions possesses a large $\zeta$ asymptotic expansion
(\cite{Abram},
10.4.18 and 10.4.63) which is obtained by keeping $\xi\neq -\frac{1}{4}$ fixed
and by expanding $f(t)$ around the extrema $t_\pm$ \rf{eq:tpm}. This is a
saddle point expansion of $A_1$ type and can be used to obtain information on
the domain where $\Phi$ is real non oscillating. Moreover we need it to recover
the expansion \rf{eq:Fguess} and to correct it (Section 3).

The residue $R(t)$ of $f(t)$ in \rf{eq:ftaylor} which is ${\cal O}((t-t_0)^4)$
and has been neglected still can be expanded systematically
\BEA
\exp\Big\{zR(t)\Big\}&\simeq&
1+\sum_{n=1}^\infty\sum_{k=2n}^\infty\,a_{n,k}\,z^n\,(t-t_0)^{2k}
\EEA
with $a_{n,k}$ polynomials in the derivatives at $t_0$. Moreover the function
\BEA
B(t)&=& t^{-1}\,(1-t)^{\gamma-1}
\EEA
can be expanded
\BEA
B(t)&=&B(t_0)\,\Big(1+\sum_{r=1}^\infty\,b_r(t-t_0)^r\Big)
\EEA
Inserting both expansions into the partition function and submitting it to the
same procedure as before we get \rf{eq:leading1F1} with $\Phi(\zeta)$ replaced
by
\BEA
\sum_{n=0}^\infty\sum_{r=0}^\infty\sum_{k=2n}^\infty\,2^{-\twrd(2k+r)}
a_{n,k}\,b_r
z^{n-\frac{1}{3}(2k+r)}\Phi^{(2k+r)}(\zeta),\quad (a_{0,0}=b_0=0)
\EEA
which is an asymptotic expansion in powers of $z^{-\frac{1}{3}}$.

\Sekt{Elementary \cats, in particular $A_2$}
Families of polynomials
\BEA
&\zeta_1 t + \zeta_2 t^2+\ldots + \zeta_{k-1} t^{k-1} \pm t^{k+1}&
\EEA
define a \cat of type $A_k$ with $\{\zeta_1,\zeta_2,\ldots
,\zeta_{k-1} \}$ as deformation parameters. Saddlepoint expansions
around deformed \cats are dealt with in the encyclopadic treatise
\cite{Arnold2}. Many questions relevant to our problem remain unanswered. In
particular we would like to know where the generalized Airy functions
$\Phi(\zeta_1,\zeta_2,\ldots
,\zeta_{k-1})$ are real positive. This domain in $I\!\!R_{k-1}$ is
bounded by a $(k-2)$-dimensional surface which can only be determined
numerically. In section 5 we will learn that this question is relevant
for even $k$ only.

In this connection the asymptotic behaviour of the functions $\Phi$ is
of interest. But even for the Pearcy function \cite{Pearcy}
\[\Phi(\zeta_1,\zeta_2)\]
which is the last one in this series carrying a name, the above series are
unknown.

Let us return to \rf{eq:partition} now and perform the angle integration
\BEA
\label{eq:partition3}\cZ_N(g)&=&
\ds\frac{\pi^{\frac{N}{2}}}{\Gamma(\Nlf)}\,\int_0^\infty\ds\frac{dt
}{t}\, \exp\Big\{\frac{N}{2}\,f(t)\Big\}
\EEA
with
\BEA \label{eq:ftdef}
f(t)&=& \log t \,-\,\sum_{k=1}^\infty\,\frac{g_k}{k}\,t^k
\EEA
We shall always normalize
\BEA
g_1&=& 1
\EEA
We want to evaluate \rf{eq:partition3} in the limit of large $N$. If
\BEA
\label{eq:fdef}&f^\prime(t_0)\,=\,0,\qquad
f^{\prime\prime}(t_0)\,\neq\,0&
\EEA
we have a Gaussian integral as leading term ($A_1$ catastrophe or Morse
singularity) implying a pure $1/N$ expansion. The $A_2$ case arises if
\BEA
\label{eq:a2cond}&f^\prime(t_0)\,=\,0,\qquad f^{\prime\prime}(t_0)\,=\,0,
\qquad f^{\prime\prime\prime}(t_0)\,\neq\,0&
\EEA
For this case to occur it is sufficient to have $g_2$ as only coupling
constant
\BEA
\label{eq:gcond}g_k &=& 0,\qquad k\ge 3
\EEA
The integral \rf{eq:partition3} can then be evaluated as a sum of two
$_1F_1$-functions, which can be treated as in the preceding section. But
the result thus obtained can be directly derived from the integral
\rf{eq:partition3} by a saddle point technique.

Solving \rf{eq:a2cond} with \rf{eq:ftdef}, \rf{eq:gcond} gives
\BEA
t_0&=& (-g_2)^{-\hlf}\,=\,\hlf\qquad (g_2<0)\\
f^{\prime\prime\prime}(t_0)&=&16
\EEA
The deformation of the \cat is achieved by one free parameter, say $\Delta$
\rf{eq:Deltad}
\BEA
\Delta&=& 1+4 \,g_2\\
f^\prime&=& (1-\Delta)^{\hlf}\,-\,1
\EEA
whereas
\BEA
t_0&=& (-g_2)^{-\hlf}
\EEA
still holds. If we expand $f(t)$ around $t_0$ at $\Delta\to 0$
\BEA
f(t)&=&f(t_0)-\hlf\Delta(t-t_0) +\frac{8}{3}(t-t_0)^3
+\,\mbox{remainder}
\EEA
and scale the integration variable
\BEA
t-t_0&=&\lambda\eta\\
\lambda&=&t_0\,\Big(\frac{2}{N}\Big)^{\frac{1}{3}}
\EEA
we obtain as leading part of the partition function
\BEA
\cZ_N(g)&\simeq&\frac{\pi^{\Nlf}}{\Gamma(\frac{N}{2})}\,
\exp\Big\{\Nlf f(t_0)\Big\}\Big(\frac{2}{N}\Big)^{\frac{1}{3}}\Phi(\zeta)
\EEA
where
\BEA
\zeta&=& \frac{1}{4}\Big(\frac{N}{2}\Big)^{\twrd}\Delta\;=\;
\Big(\Nlf\Big)^{\twrd} (g_2-g_2^*),\qquad g_2^*\,=\,-\frac{1}{4}
\EEA
and
\BEA
\Phi(\zeta)&=&\int\limits_{C^\prime}\,d\eta e^{-\zeta\eta +\frac{1}{3}\eta^3}
\EEA
The contour $C^\prime$ must be chosen such that under replacement of
\BEA
\eta&\to&-\eta\nonumber
\EEA
$\Phi(\zeta)$ becomes identical with \rf{eq:wefind}.

In the case of the Airy functions asymptotic expansions for large $\zeta$ need
not be obtained from a saddle point expansion but \cite{Abram}, 10.4.63 tells
us that
\BEA
\mbox{Bi}(\zeta)&\simeq& \pi^{-\hlf}\zeta^{-\frac{1}{4}} e^z
\sum_{k=0}^\infty\,c_k\,z^{-k}\\
z&=&\twrd\,\zeta^{\tlf}\\
c_k&=& \ds\frac{\Gamma(3k+\hlf)}{54^k k!\Gamma(k+\hlf)}
\EEA
which agrees with the saddle point expansion.

If we now take the logarithm we obtain
\BEA
\log \Phi(\zeta)&=& \hlf\log\pi -\frac{1}{4}\log\zeta
+\twrd\zeta^{\tlf} +c_1\tlf\zeta^{-\tlf} + (c_2-\hlf
c_1^2)\frac{9}{4}\zeta^{-3}+\ldots
\EEA
Comparison with \rf{eq:Fdef}
\BEA
F_N(g_2)&=& \hlf -\hlf f(t_0) + \frac{1}{3 N}\log \frac{N}{2}
-\frac{1}{N}\log\Phi
\EEA
and \rf{eq:Fguess} allows us to identify the coefficients $a_h$, namely
\BEA
-a_0&=&+\frac{1}{3}\nonumber\\
-a_1&=& \mbox{arbitrary}\nonumber\\
-a_2&=& 3\,c_1\nonumber\\
-a_3&=& 9\,(c_2 -\hlf c_1^2)
\EEA
Moreover the singular term
\BEA
+\frac{1}{4N}\log\zeta &=& \frac{1}{6N}\log\Nlf (g_2-g_*)^\tlf
\EEA
implies \rf{eq:a1eq}.

The partition function diverges at the critical value $g_c = g_2^*
=-\frac{1}{4}$. We shall show in the subsequent section that adding a term to
$f(t)$  \rf{eq:ftdef}, \rf{eq:gcond}
\BEA
-\frac{1}{3}g_3 t^3 ,&& g_3>0,\quad\mbox{small}
\EEA
is the most elegant way to come around this problem.

\Sekt{Deformation of an $A_3$ \cat into curves of $A_2$ \cats}
Let
\BEA \label{eq:ftdeform}
f(t)&=& \log t -t -\hlf g_2 t -\frac{1}{3} g_3 t^3
\EEA
where $g_3$ is a free parameter but we assume still that
\BEA
f^{\prime\prime\prime}(t_0)&\neq& 0
\EEA
then $g_3$ defines a curve of $A_2$ catastrophes. This curve has two real
analytic
branches intersecting at the $A_3$ \cat (Figs 1 and 2), where
\BEA
f^{\prime\prime\prime}(t_0)&=&0
\EEA
In fact
\BEA
f^{\prime}(t_0) &=& f^{\prime\prime}(t_0)\,=\,0
\EEA
can be solved in the form (see \cite{Hig}, eqns (78),(79))
\BEA
t_{0,\pm}&=& -\frac{1}{g_2}\Big[ 1\pm (1+3g_2)^\hlf\Big]
\EEA
and
\BEA
-27g_{3,\pm}&=& 2+ 9g_2 \mp 2(1+3g_2)^\tlf
\EEA
These are the curves drawn in Figs. 1 and 2 respectively.

The partition function is convergent with \rf{eq:ftdeform} if $g_3>0$. We
mentioned already in Section 3 that an arbitrary small positive $g_3$ can be
used to give a well defined meaning to the $t_-$-branch (at the dot). If we
move
from this point towards the $A_3$ \cat we have
\BEA
\begin{array}{rcccl}
0&\le&27 g_3 &<& 1\end{array}\\
\begin{array}{rcccl}
-\frac{1}{3} &<g_2&\le & -\frac{1}{4}\end{array}
\EEA
and remain continuously connected to the cuspoidal $A_2$ catastrophe. This
makes sense if we
can steer the parameters $g_2$, $g_3$ at will. What if the system is such that
it can adjust the parameter $g_3$ freely for fixed $g_2$ ? Then it could jump
(first order transition) to the $t_+$-branch eventually. However, since at $g_2
=-\frac{1}{4}$
\BEA
f(t_{0,+})-f(t_{0,-})\bigg|_{g_2=-\frac{1}{4}}&=&\log 3
-\frac{4}{3}\quad(=\;-0.2347)
\EEA
the $t_-$-branch is stable.

We consider now $g_2$ and
\BEA
\Delta&=& -2 f^\prime (t_0)
\EEA
as deformation parameters. Then
\BEA
t_{0,\pm}^{-1}&=& \frac{1}{3}\bigg[(1-\frac{\Delta}{2})\mp \Big(
(1-\frac{\Delta}{2})^2 +3g_2\Big)^\hlf\bigg]
\EEA
Our formulas are applicable in the case
\BEA
g_2<0,&&\Delta<2
\EEA
and
\BEA \label{eq:Delgcond}
\Big(1-\frac{\Delta}{2}\Big)^2 +3g_2 &>&0
\EEA
Setting \rf{eq:Delgcond} equal zero gives the $A_3$ \cat since
\BEA
f^{\prime\prime\prime}(t_{0,\pm})&=& \mp 2 t_{0,\pm}^{-2}
\bigg[\Big(1-\frac{\Delta}{2}\Big)^2 +3g_2\bigg]^\hlf
\EEA
In order to avoid oscillating Airy functions which would lead to complex free
energies we must moreover have
\BEA \label{eq:sgnconstraint}
\mbox{sign} \Big\{\Delta f^{\prime\prime\prime}(t_{0,\pm})\Big\}&=& +1
\EEA

Now we scale
\BEA
t-t_0 &=& \lambda\eta
\EEA
in
\BEA
f(t)&=& f(t_0)-\hlf\Delta (t-t_0) +\frac{1}{6}f^{\prime\prime\prime}(t_0)
 (t-t_0)^3 +{\cal O}\Big( (t-t_0)^4\Big)
\EEA
so that
\BEA
\Nlf \frac{1}{6} \lambda^3\Big| f^{\prime\prime\prime}(t_0)\Big|&=&
\frac{1}{3}
\EEA
or
\BEA
\lambda&=&\left( \ds\frac{4}{N | f^{\prime\prime\prime}(t_0)
|}\right)^{\frac{1}{3}}
\EEA
The contribution of either branch to the partition function is
(leading term only)
\BEA
\cZ_N(g_2,g_3)_\pm&\simeq&
\frac{\pi^{\Nlf}}{\Gamma(\Nlf)}\exp\Big\{\Nlf
f(t_{0,\pm})\Big\}\,t_{0,\pm}^{-1}
\,\left(\ds\frac{4}{N | f^{\prime\prime\prime}(t_{0,\pm})
|}\right)^{\frac{1}{3}}\Phi_\pm(\zeta_\pm)
\EEA
where
\BEA
\zeta_\pm&=& \frac{1}{4} N^{\twrd} |\Delta|\left|
\ds\frac{4}{f^{\prime\prime\prime}(t_{0,\pm})}\right|^{\frac{1}{3}}
\EEA
and
\BEA \label{eq:thecontours}
\Phi_\pm(\zeta_\pm)&=&\int_{C_\pm}
d\eta\,\exp\Big\{\pm\zeta\eta\mp\frac{1}{3}\eta^3\Big\}
\EEA
$C_\pm$ are the contours obtained by a minimal deformation of the positively
orientated real axis that makes the integrals convergent. We obtain
\BEA
\Phi_\pm(\zeta)&=&\pi\mbox{Bi}(\zeta)
\EEA
in either case. The constraint \rf{eq:sgnconstraint} renders $\zeta$ positive
in
either case.

\Sekt{The general $A_m$ (cuspoidal) \cat}
We consider now the case
\BEA
&g_1\,=\,1,\quad g_k\,=\,0 ,\quad k\ge m+1&
\EEA
In this case we have
\BEA
f^{(m+1)}(t)&=&(-1)^m\frac{m!}{t^{m+1}}\;\neq\;0\quad\forall t
\EEA
The $A_m$ \cat occurs if
\BEA \label{eq:occurs}
f^{(k)}(t_0)&=&0\quad\forall\; 1\le k \le m
\EEA
This leads to the equations
\BEA \label{eq:system}
\ds\frac{(-1)^{k-1}}{t_0^k}\,-\,\sum_{l=k}^m\,g_l {l-1 \choose k-1} t_0^{l-k}
&=& 0
\EEA
This system can be solved in an elementary fashion for $t_0$ and $\{g_l\}_2^m$

First we set
\BEA \label{eq:firstweset}
k=m:\quad \ds\frac{(-1)^{m-1}}{t_0^m}&=& g_m
\EEA
which entails
\BEA \label{eq:entails}
\mbox{sign}\; g_m &=& (-1)^{m-1}
\EEA
since $t_0$ must be positive
\BEA
t_0&=&\left| g_m\right|^{-\frac{1}{m}}
\EEA
Convergence of the partition function necessitates
\BEA
\mbox{sign}\; g_m &=& +1
\EEA
which is compatible with \rf{eq:entails} only if m is odd. For even $m$ we will
employ the procedure discussed in the preceding section: We add a term
\BEA
&-\frac{1}{m+1}\, g_{m+1}\, t^{m+1},\quad g_{m+1}\searrow 0
\EEA
to the action $f(t)$.

An intermediary step in solving the system \rf{eq:system} is
\BEA \label{eq:intermed}
g_k&=& (-1)^{m-k}\, {m \choose k} \,t_0^{m-k}\, g_m,\qquad (1\le k\le m)
\EEA
which for $k=m$ is trivial. For $k=1$ we make use of $g_1=1$ to obtain $t_0=m$.
In proving \rf{eq:intermed} we need the identity
\BEA \label{eq:identity}
\sum_{l=k}^m\,(-1)^{l-k}\,{l-1 \choose k-1} {m \choose l} &=& 1
\EEA
Denote the l.h.s. of \rf{eq:identity} by $P_k^{(m)}$. Then
\BEA
P_k^{(m)}-P_{k+1}^{(m)}&=& \sum_l (-1)^{l-k}\,{m \choose l}
\Big[ {l-1 \choose k-1} + {l-1 \choose k}\Big]
\EEA
which by Pascal's identity gives
\BEA
&=& {m \choose k} \sum_l (-1)^{l-k} {m-k \choose l-k} \nonumber\\
&=& \delta_{mk}
\EEA
Since $P_m^{(m)}=1$ \rf{eq:identity} follows. Inserting \rf{eq:firstweset} into
\rf{eq:intermed} we have finally
\BEA
g_k &=& (-1)^{k-1} {m \choose k} \,m^{-k}
\EEA

Now we deform this \cat by
\BEA \label{eq:deformthiscat}
g_k &=& (-1)^{k-1} {m \choose k} \,m^{-k} +\tau_k,\qquad (2\le k\le m)\\
t_0&=& m+\tau_0
\EEA
Referring to translational invariance in $t$ we postulate
\BEA \label{eq:postulate}
f^{(m)}(t_0) &=& 0
\EEA
so that
\BEA
f(t) &=& \sum_{l=0}^{m-1} \frac{(t-t_0)^l}{l!}\, f^{(l)}(t_0)\,+\,
\frac{(t-t_0)^{m+1}}{(m+1)!}\,f^{(m+1)}(t_0)\,+\, {\cal O}\Big(
(t-t_0)^{m+2}\Big)
\EEA
We expand $f^{(l)}(t_0)$ linearly in all $\tau_0$ and $\tau_k$, $2\le k \le m$.

First we notice that
\BEA
\frac{\partial}{\partial \tau_0} f^{(l)}(m+\tau_0)\bigg|_{\tau_0 =
\tau_k = 0\,\forall k}&=&0,\quad 0\le l\le m-1
\EEA
since
\BEA
f^{(l+1)}(m)&=&0
\EEA
from \rf{eq:occurs}. So there remains $(\tau_1 = 0)$
\BEA \label{eq:remains}
f^{(l)} (m+\tau_0)&=& -(l-1)!\sum_{k=l}^m {k-1 \choose l-1} m^{k-l}
\tau_k+\,\mbox{quadratic terms in $\{\tau\}$}
\EEA
Next comes the scaling procedure
\BEA
t-t_0&=&\lambda\eta
\EEA
so that
\BEA
\Nlf\,\frac{\lambda^{m+1}}{(m+1)!}\,\Big|f^{(m+1)}(m)\Big|&=&\frac{1}{m+1}
\EEA
leading to
\BEA
\lambda&=& m\Big(\frac{2}{N}\Big)^{\frac{1}{m+1}}
\EEA
Then introduce the scaling variables
\BEA
\zeta_l &=& \Nlf \frac{\lambda^l}{l!}\,f^{(l)}(m+\tau_0)
,\qquad (1\le l \le m-1)
\EEA
which are kept ${\cal O}(1)$ at the transition point by definition. It follows
\BEA \label{eq:transitionpoint}
f^{(l)}(m+\tau_0) &=& A_l\, \Big(\Nlf\Big)^{-\sigma_l}
\EEA
where
\BEA
{\sigma}_l &=& 1 - \frac{l}{m+1}
\EEA
and
\BEA \label{eq:AlO1}
A_l &=& \frac{l!}{m^l}\,\zeta_l \;=\;{\cal O}(1)
\EEA
so we have simple scaling of the derivatives implying that the coupling
constants
\BEA
&\tau_2,\,\tau_3,\,\ldots ,\,\tau_{m-1},\,\tau_m
\EEA
scale along algebraic curves with $\Nlf$ as single parameter,
whereas $\tau_0$ is coupled to $\tau_m$ by
\BEA
\tau_0&=& (-1)^m m^m \tau_m
\EEA
as follows from \rf{eq:postulate}. In order to solve \rf{eq:remains}  for the
$\tau_k$ we denote
\BEA \label{eq:denoteNlk}
\cN_{lk}&=& {k-1 \choose l-1}\,m^{k-l}
\EEA
Then the matrix inverse is
\BEA
\cN_{kl}^{-1}&=& (-1)^{l-k}\,\left[ {l-1 \choose k-1} -{m-1 \choose
k-1}\right]\,m^{l-k},
\qquad\left\{\begin{array}{ll}1\le l\le m-1\\2\le k\le m \end{array}\right\}
\EEA
It follows
\BEA
\tau_k &=& -\sum_{l=1}^{m-1}
\cN_{kl}^{-1}\frac{A_l}{(l-1)!}\,\Big(\Nlf\Big)^{-\sigma_l}
\EEA
Note that $\tau_1$ vanishes automatically.

The contribution of this saddle point to the partition function is
\BEA \label{eq:leading}
\cZ_N(g)\Big|_{A_m}&=&
\frac{\pi^{\Nlf}}{\Gamma(\Nlf)}\exp\Big\{\Nlf f(m)\Big\}\,
\Big(
\frac{2}{N}\Big)^{\frac{1}{m+1}}\,\Phi(\zeta_1,\zeta_2,\ldots,\zeta_{m-1})
\EEA
as leading term where
\BEA \label{eq:leadingterm}
\Phi(\zeta_1,\zeta_2,\ldots,\zeta_{m-1}) &=& \int\limits_{C^{(m)}}
d\eta\,\exp\left\{\sum\limits_{k=1}^{m-1}\zeta_k\eta^k +
(-1)^m\frac{\eta^{m+1}}{m+1} \right\}
\EEA
For odd $m$ we identify $C^{(m)}$ with the positive real axis. For even $m$ we
define
\BEA
C^{(m)}&=& - C_{\hlf} + \hlf\left\{ C_{\frac{1}{2(m+1)}}
+C_{\frac{-1}{2(m+1)}}\right\}
\EEA

What are the conditions for asymptotic behaviour of $\Phi$ to be nonoscillating
for large $\{\zeta_k\}$ ? For
$m$ odd there is no problem since
\BEA
m\;\mbox{odd} &:& \Phi(\zeta)\,>\,0
\EEA
by our definition. For even $m$ only the case $m=2$ has been dealt with
already, e.g. by condition
\rf{eq:sgnconstraint} that leads to the correlation of signs in
\rf{eq:thecontours}. For $m\ge 4$ studying the
contributions of all subordinate catastrophes
\[A_n\,,\,n\,<\,m\]
is a complicated algebraic task. An approach giving an insight into the case
$m=4$ is presented in the Appendix.
\Sekt{Differential equations and the renormalization equation}
The function $\Phi(\zeta_1,\zeta_2,\ldots,\zeta_{m-1})$ \rf{eq:leadingterm}
satisfies the following set of linear differential equations
\BEA
\frac{\partial\Phi}{\partial\zeta_k} &=&
\frac{\partial^k\Phi}{\partial\zeta_1^k}
,\qquad k\in\{1,2,\ldots m-1\}
\EEA
and
\BEA
(-1)^m\, \frac{\partial^m}{\partial\zeta_1^m}\Phi\,+\,\sum_{k=1}^{m-1}
k\zeta_k\frac{\partial}{\partial\zeta_{k-1}}\Phi &=& 0
\EEA
In turn this system of $m-1$ equations determines $\Phi$ to lie in the
$m$-dimensional space of integrals \rf{eq:leading} with admissable contours
$C^{(m)}$.

Let us denote
\BEA
F(\zeta)&=& \log \Phi (\zeta)
\EEA
Then $F(\zeta)$ satisfies a system of nonlinear differential equations, e.g.
for
$m=2$
\BEA
F^{\prime\prime} + (F^\prime)^2&=&\zeta
\EEA
which is the Airy differential equation in logarithmic camouflage.

Independently of these differential
equations $F$ satisfies the renormalization group equation
\BEA \label{eq:rengroup}
\left( N\frac{\partial}{\partial N} -\sum_{k=2}^m\,\beta_k (\tau)
\frac{\partial}{\partial\tau_k}\right) F(\zeta)&=& 0
\EEA
where we used the deviations $\tau_k$ of the coupling constants $g_k$ off their
critical values \rf{eq:deformthiscat}. Knowledge of the variables $\{\zeta_l\}$
as functions of $N$ and $\{\tau_k\}$ allows us to determine $\beta_k(\tau)$
(insert \rf{eq:remains} and \rf{eq:AlO1} into \rf{eq:transitionpoint})
\BEA
\beta_k(\tau)&=&
\sum_{\tau_l}\,\frac{\partial\beta_k}{\partial\tau_l}\bigg|_{\tau=0}\,\tau_l +
\cO (\tau^2)
\EEA
We find
\BEA
N\frac{\partial}{\partial N} F&=&
\sum_l\,\sigma_l\zeta_l\frac{\partial}{\partial\zeta_l} F\\
\frac{\partial}{\partial\tau_k} F&=& \sum_l
\frac{\partial\zeta_l}{\partial\tau_k}
\frac{\partial}{\partial\zeta_l} F
\EEA
and from \rf{eq:rengroup}, \rf{eq:denoteNlk}
\BEA
\sum_l \left\{ \sigma_l\sum_k \cN_{lk}\tau_k\,-\,\sum_k \cN_{lk}\beta_k(\tau)
+{\cal O}(\tau^2) \right\}\,\frac{\partial F}{\partial \zeta_l} &=& 0
\EEA
Setting each coefficient of $\frac{\partial F}{\partial \zeta_l}$ equal to zero
gives
\BEA
\sigma_l \cN_{lk} &=& \sum_r \cN_{lr} \xi_{rk}
\EEA
with the susceptibility matrix
\BEA
\xi_{rk}&=&
\frac{\partial\beta_r(\tau)}{\partial\tau_k}\bigg|_{\tau_l=0\,\forall
l}
\EEA
So the matrix $\cN_{lr}$ diagonalizes this susceptibility matrix and
$\{\sigma_l\}$ are its eigenvalues. In our inverse approach we can calculate
$\xi$
by
\BEA
\xi_{rk} &=& \sum_l\cN_{rl}^{-1}\sigma_l\cN_{lk}
\EEA
The sum can be performed and gives
\BEA
\beta_k(\tau)&=& (-1)^k {m-1 \choose k-1} \frac{m^{2-k}}{m+1}\tau_2
+\frac{m+1-k}{m+1}\tau_k + \frac{mk}{m+1}(1-\delta_{km})\tau_{k+1} +\cO
(\tau^2)
\qquad\Balk{1}\\
&&\qquad\qquad\qquad\qquad\qquad\qquad\qquad\qquad\qquad\qquad\qquad\qquad\qquad(2\le k\le m)\nonumber
\EEA
\renewcommand{\theequation}{A.\arabic{equation}}
\renewcommand{\thesection}{Appendix:}
\Sekt{Generalized Airy functions for $A_m$, $m$ even}
The asymptotic expansions of generalized Airy functions for large arguments as
defined in \rf{eq:leadingterm}
are treated with the same saddle point techniques which produced them. An Airy
function for a \cat $A_m$
obtains contributions of all \cats $A_l$, $1<l<m$.

We restrict the arguments by
\BEA
\zeta_i\,=\,0,\qquad k<i\le m-1
\EEA
and consider corresponding reduced phase functions
\BEA
f_k(\eta)&=&\sum\limits_{i=1}^k \zeta_i\eta^i\,+\, \frac{1}{m+1}\,\eta^{m+1}
\EEA
We let
\BEA
\zeta_1\,=\,\epsilon | \zeta_1 |,\qquad | \zeta_1 |
\to\infty,\qquad\epsilon^2\,=\,1
\EEA
and couple the remaining variables to $| \zeta_1 |$
\BEA
\zeta_i&=&\alpha \, | \zeta_1 |^{\frac{m+1-i}{m}},\qquad 2\le i <k \\
\eta_0&=&\omega \, | \zeta_1 |^{\frac{1}{m}}
\EEA
where $\eta_0$ is the position of the $A_l$ catastrophe,
so that
\[\{\alpha_2,\alpha_3,\ldots,\omega\}\,\in\,I\!\!R^k\]
is fixed in the limiting procedure. Then
\BEA
f_k(\eta_0)&=& \Psi_0(\alpha_2,\alpha_3,\ldots,\omega)\,| \zeta_1
|^{\frac{m+1}{m}}\\
\Psi_0(\alpha_2,\alpha_3,\ldots,\omega)&=& \epsilon\,\omega +
\sum\limits_{l=2}^k\,\alpha_l\omega^l +
\frac{1}{m+1}\omega^{m+1}
\EEA
We introduce the shorthand
\BEA
\Psi_l(\alpha_2,\alpha_3,\ldots,\omega) &=& \frac{\partial^l}{\partial\omega^l}
\Psi_0(\alpha_2,\alpha_3,\ldots,\omega)
\EEA
Then the $A_1$ \cats appear at
\BEA
\Psi_1(\alpha_2,\alpha_3,\ldots,\omega) &=& 0
\EEA
In the case $k=1$ this is
\BEA
\Psi_1(\omega)&=& \epsilon + \omega^m \;=\; 0
\EEA
and has $m$ solutions $\omega$. Only that solution is relevant in our context
for which
\[\Re \left( \Psi_0\Big|_{\Psi_1=0}\right)\]
is maximal. For a real nonoscillating asymptotic expansion we need moreover
\BEA \label{eq:a11}
\Psi_0\;\mbox{real},\qquad \Psi_2<0
\EEA
For $k=1$ this is possible only if
\BEA \label{eq:a12}
\epsilon\,=\,-1,\qquad\omega\,=\,-1
\EEA

Let us consider the case $k=2$ in more detail and concentrate on the
case \rf{eq:a11}. Then there are two possibilitues
\BEA \label{eq:a13}
\epsilon\,=\,-1,\qquad-\infty<\omega < 0,\qquad
\alpha_2\,=\,\frac{1-\omega^m}{2\omega}
\EEA
or
\BEA\label{eq:a14}
\epsilon\,=\,+1,\qquad 0<\omega < (m-1)^{-\frac{1}{m}},\qquad
\alpha_2\,=\,-\frac{1+\omega^m}{2\omega}
\EEA
The case \rf{eq:a12} is contained in \rf{eq:a13} but \rf{eq:a14} is a new
branch. At the end of this branch we
have
\BEA
\omega&=&(m-1)^{-\frac{1}{m}}
\EEA
and an $A_2$ catastrophe. Denoting
\BEA
\chi_k&=& \Psi_k\Big |_{\Psi_1=0}
\EEA
the $A_1$ \cats contribute as leading term
\BEA
\Phi(\zeta_1,\zeta_2,\ldots,\zeta_k,0,\ldots)&\simeq& e^{\ds\chi_0
|\zeta_1|^{\frac{m+1}{m}}}\left [
\ds\frac{2\pi}{|\chi_2 | \,|\zeta_1|^{\frac{m-1}{m}}}\right ]^{\hlf}
\EEA
Now to the $A_2$ catastrophes. They show up first at $k=2$ as we just found,
and are located at
\BEA \label{eq:a18}
\Psi_1&=&0,\\ \label{eq:a19}
\Psi_2&=&0
\EEA
For $k=2$ these conditions are solved by
\BEA
\epsilon&=& (m-1)\,\omega^m\\ \label{eq:a21}
\alpha_2&=&-\hlf m\,\omega^{m-1}
\EEA
We deform the condition \rf{eq:a18} by a new parameter and use the standard
scaling technique
\BEA
\Psi_1&=&\Delta_1\\
\epsilon&=&(m-1)\,\omega^m\,+\,\Delta_1
\EEA
but \rf{eq:a19}, \rf{eq:a21} is maintained. With
\BEA
\lambda&=& \left [\ds\frac{2}{|\chi_3 | \,|\zeta_1|^{\frac{m-2}{m}}}\right
]^{\trd}
\EEA
and
\BEA \label{eq:a25}
z&=&\mbox{sign}\,\chi_3\,\Delta_1\,\lambda\,|\zeta_1 |
\EEA
we obtain the leading term of the asymptotic expansion $(k\ge 2)$
\BEA
\Phi(\zeta_1,\zeta_2,\ldots,\zeta_k,0,\ldots)&\simeq & e^{\ds\chi_0
|\zeta_1|^{\frac{m+1}{m}}}\,
\left[ \ds\frac{2}{| \chi_3 | \,|\zeta_1
|^{\frac{m-2}{m}}}\right]^{\trd}\,\Phi(z)
\EEA
if $\Delta_1$ approaches zero in such a fashion that $z$ \rf{eq:a25} remains
fixed. Moreover $\chi_k$ is now
obtained from $\Psi_k$ by restriction to \rf{eq:a18} and \rf{eq:a19}.

It is obvious that this procedure can be extended to all $A_l$ and $k$.

\end{document}